\documentclass[prd,nofootinbib,showpacs,showkeys,twocolumn]{revtex4}

\usepackage{graphicx}
\usepackage{amsmath}
\usepackage{amssymb}
\usepackage{amsfonts}
\usepackage{latexsym}
\usepackage{mathrsfs}
\usepackage{color}
\usepackage{slashed}
\usepackage{float}
\usepackage{feyn}
\usepackage[all]{xy}
\usepackage{feyn}
\usepackage{hyperref}
\usepackage{dsfont}
\usepackage[justification=centerlast,singlelinecheck=false,font=small]{caption,subfig} 
\usepackage{feynmp}
\definecolor{blue}{rgb}{0,0,0.5}
\definecolor{lightblue}{rgb}{0,0,1}
\definecolor{red}{rgb}{0.5,0,0}
\definecolor{lightred}{rgb}{1,0.5,0}
\definecolor{green}{rgb}{0,0.5,0}
\definecolor{darkgreen}{rgb}{0.0,0.3,0.0}
\definecolor{orange}{rgb}{1,0.4,0}
\definecolor{grey}{rgb}{0.5,0.5,0.5}
 
\providecommand{\eqn}{Eq.~}

\newcommand{\printifnonempty}[2]{\if\relax\detokenize{#1}\relax\else#2\fi}
\newcommand{\alter}[5]{%
  \long\def\temp{#3}%
  \long\def\accept{#5}%
  \ifx\temp\accept
    {#1}
  \else
    {\textcolor{#4}{\printifnonempty{#1}{{#1}}}%
    \textcolor{grey}{\printifnonempty{#2}{(#2)}}%
    \textcolor{#4}{\printifnonempty{#3}{{[#3]}}}}%
  \fi
}

\global\long\def\m{\mu}

\global\long\def\n{\nu}
\global\long\def\d{\partial}
\global\long\def\na{\nabla}
\global\long\def\l{\lambda}
\global\long\def\v{\delta}
\global\long\def\G{\Gamma}
\global\long\def\r{\rho}

\begin{document}

\pacs{miow}

\title{Regularization of Vacuum Fluctuations and Frame Dependence}

\author{Juri Smirnov}
\email[\,]{juri.smirnov@mpi-hd.mpg.de}

\affiliation{
Max-Planck-Institut f\"ur Kernphysik, Saupfercheckweg 1, 69117 Heidelberg, Germany}

\begin{abstract}
We discuss the regularization of vacuum fluctuations in a gravitational background. It is shown that general covariance is broken even by a 4-momentum cut-off, consistent with Lorentzian symmetry. It is pointed out that general covariance is a protective symmetry for the vacuum energy from power divergences. 
\end{abstract}

\maketitle

\section{\label{sec:Introduction}Introduction}

In quantum field theories contact diagrams diverge. This
problem is addressed  by the technique of regularization and renormalization.
Mathematically the divergences arise from multiplication of distributions which is
a priori an undefined operation.
This problem can be addressed by regarding distributions
as functions almost everywhere and  subtracting the problematic point. The defined
quantity has to be continued again to the whole space. This is done
by adding a counterterm corresponding to the graph where the divergent loop
is contracted to a point.
This procedure is equally valid on curved space-time backgrounds as shown in \cite{Prange:1997iy,Fredenhagen}.
The important fact is that the coefficient of the counterterm is chosen {}``by hand'' or
fixed by experiment i.e. the field theory has no prediction at all
for this value. We see that the very nature of renormalization
is such that a prediction is impossible. Particularly the statement of naturalness in this sense is meaningless.

In the process of renormalization a technique is needed to quantify
the degree of divergence of a diagram, therefore a regulator is introduced which is viewed as
a mathematical tool and thus has to be removed at the end of the procedure.
We argue that unless this regulator is consistent with the symmetries of the theory counterterms have to be employed which restore this symmetry and thus 
the resulting values can not be viewed as predictions of physical quantities. We generalize the three dimensional cut-off regularization used in \cite{Casadio:2013uia, Mangano:2010hw} to four dimensions and find that even though it preserves Lorentz symmetry, general covariance is violated.
We find that in fact general covariance acts as a protective symmetry for the vacuum energy in the same way as conformal symmetry is proposed to protect the Higgs mass \cite{Bardeen:1995kv}.

\section{\label{sec:VacFluc} Vacuum fluctuations in semiclassical gravity }

Since the observation of Casimir forces \cite{Lamoreaux} and recently even the dynamical Casimir effect \cite{DC} it is clear that the vacuum
fluctuations of fields are not just mathematical peculiarities
of the theory but physical reality. Thus we expect an impact of this
vacuum energies on gravity. Since it is the energy momentum tensor
which shapes our space-time we will make the attempt to calculate
the contribution of vacuum fluctuations to this quantity. Since we can not solve the full dynamics of the problem the average effect 
given by the vacuum expectation value of the energy-momentum tensor of the vacuum is computed.
Two major consistency criteria need to be obeyed.

\subsection*{ Choice of the energy scale}
The first criterion is the existence of a self consistent asymptotically flat space-time solution. 
We argue that a non zero vacuum expectation value of the asymptotically flat space-time leads to a contradiction.
Assume there exists a asymptotically flat solution
of a given Einstein equation. If the vacuum tensor of this space-time would not vanish, its energy
distribution had the same symmetry as itself and hence would be space-time
uniform. If this energy would couple to gravity by means of entering
the energy momentum tensor in Einstein's equation, the space-time could
not be an asymptotically flat solution of the Einstein equation. Thus we arrive at a contradiction. 

Also other approaches exist which would support the disappearance of
space time uniform vacuum energy, for instance the idea of degravitation
due to a finite graviton width  \cite{Dvali}, which suggest that a space-time
uniform source would not contribute to the energy momentum tensor
on cosmological scales.

Thus we define a normalization of the vacuum energy momentum tensor in the following way:

\begin{align}
\langle0|T{}_{\m\n}|0\rangle_{\text{vac}}=\langle0|T{}_{\m\n}|0\rangle_{\text{FRW bare}}-\langle0|T{}_{\m\n}|0\rangle_{\text{H $\rightarrow$ 0}}.
\end{align}

This normalization is by no means in contradiction to usual quantum field theories, as in flat space-time the zero point energy is simply irrelevant.  The normalization condition corresponds to subtracting the the zero order contribution in the adiabatic subtraction \cite{Fulling:1974pu} and will allow us to study second order contributions to vacuum energy.

\subsection*{General covariance}

The second consistency condition is that the resulting equations obey the principle of general covariance. This principle is used to construct
the action of general relativity and it dictates that the Bianchi
identity has to hold even off shell, it reads 

\begin{align}
\na^{\m}G_{\m\n}=\na^{\m}(R_{\m\n}-\frac{1}{2}g_{\m\n}R)=0.
\end{align}

Thus, if Einstein's equation holds general covariance dictates $\na^{\m}T_{\m\n}^{\text{total}}=0$.
In the case of an FRW background symmetry this equation is equivalent
to $\na^{\m}T_{\m\n}^{\text{total}}=\dot{\rho}+3H\left(\r+p\right)=0$. 

\subsection{The minimal set up}

Let us consider the minimal set up, where a massless scalar field
$\phi$ is minimally coupled to gravity, hence the Lagrangian we are
working with is \cite{Maggiore,Birrell:1982ix}:

\begin{align}
\mathcal{L}=\sqrt{-g}\left\{ \frac{1}{16\pi G_{N}}\, R-\frac{1}{2}\d_{\m}\phi\d^{\m}\phi\right\} .
\end{align}

We will use a semi classical approach to this problem, where a gravitational
classical background is assumed to exist and the matter and metric
perturbation fields are considered to be quantized on that background.
We will demonstrate now the scalar field quantization. For the definition
of quantum field operators we need a complete orthonormal set of mode
solutions. To this end we define a scalar product for the fields by
taking a Cauchy hyper-surface $\Sigma$ 
on the pseudo-riemanian manifold $\mathcal{M}$ defined by
an orthogonal future directed flow $u^{\m}$. This enables us to define
a scalar product

\begin{align}
\langle\phi|\psi\rangle=-i\int_{\Sigma}\sqrt{h}(\phi\d_{\m}\psi*-(\d_{\m}\phi)\psi*)u^{\m}d\Sigma,
\end{align}

here $h$ is the metric induced on $\Sigma$. 

We assume the existence of the set of solutions orthonormal in the
sense of this scalar product and define raising and lowering operators
with respect to them. This means that we have selected a Fock space
construction and have chosen a vacuum state. This construction is
a direct consequence of the time-space split discussed above.

The scalar field operator can be written as a mode expansion with
the raising and lowering operators

\begin{align}
\phi(x)=\int\frac{d^{3}k}{(2\pi)^{3}2k}\left[a_{k}\phi_{k}(t)\, e^{ikx}+h.c.\right].
\end{align}

Varying the action w.r.t. $\phi$ we obtain the equation of motion
the field has to obey. The wave equation reads in the FRW background
as:

\begin{align}
\Box\phi(x)=\na_{\m}g^{\m\n}\d_{\n}\phi=-\d_{0}^{2}\phi-\G_{0\l}^{\l}\d_{0}\phi+\frac{1}{a^{2}}\vec{\d}_{i}^{2}\phi=0.
\end{align}

Which for the individual modes implies

\begin{align}
& \d_{0}^{2}\phi_{k}+3\frac{\dot{a}}{a}\d_{0}\phi_{k}+\frac{1}{a^{2}}k^{2}\phi_{k}=0\quad\Leftrightarrow\quad\phi''_{k}+2H\phi'_{k}+k^{2}\phi_{k}=0 \\ \nonumber  & \text{with}\;(\cdot)':=\frac{\d}{\d\eta}\;\text{and}\;\eta=\int\frac{dt}{a}.
\end{align}

Defining $(\cdot)':=\frac{\d}{\d\eta}\;\text{with}\;\eta=\int\frac{dt}{a}$ and  $\psi_{k}:=\phi_{k}/a$ the equation of motion for the individual modes reads

\begin{align}
\psi_{k}''+\left(k^{2}-\frac{a''}{a}\right)\psi_{k}=0.
\end{align}

Various solutions to this equation are known which form depends on
the background type. Choosing a de Sitter background where $a=-1/(H\eta)$
the solution is \cite{Maggiore}

\begin{align}
\phi_{k}(\eta)=\frac{1}{a}\left(1-\frac{i}{k\eta}\right)e^{-ik\eta}=e^{-ik\eta}\left(\frac{i}{k}-\eta\right)H.
\end{align}

\subsection{The energy momentum tensor}

In this set up, however, the particular dynamics of the field is not
of interest but its average effect on the space time. Therefore,
we seek an expression for the vacuum expectation value of the energy momentum of the field

\begin{align}
T_{\m\n}=\d_{\m}\phi\d_{\n}\phi-\frac{1}{2}g_{\m\n}\d_{\l}\phi\d^{\l}\phi.
\end{align}

To define the energy density and pressure consistent with our time and space split we use the frame field 
$u^\mu$ to define the time direction. Note that it fulfils the normalization condition $u_\mu u^\mu = -N$ with $N$ being the lapse function to stay general. Hence the orthogonal projector to this direction is defined by 
\begin{align}
h^{\mu\nu} = u^\mu u^\nu + g^{\mu \nu}.
\end{align} 
And the induced metric $h_{ij}$ on the spatial hypersurface $\Sigma$ is the pullback of it, with $X$ the coordinates on the 4 manifold and $x$ coordinates on the 3 manifold
\begin{align}
h_{ij} = h_{\mu\nu}\frac{\partial X^\mu}{\partial x^i} \frac{\partial X^\nu}{\partial x^j} = : h_{\mu\nu} X^\mu_i X^\nu_j .
\end{align} 

We can now define the energy and pressure with respect to this direction by 
\begin{align}
\rho =  u^\mu u^\nu T_{\mu \nu}  \text{  and  } 3 p = h^{\mu i}h^{\nu}_i T_{\mu \nu}.
\end{align} 
Using that $X^\mu_i u_\mu =0$ which is just the orthogonality condition, $\partial_\mu X^\mu_i=\partial_i$ and $h_i^i=3$ we get

 \begin{align}
\rho =  N \int\frac{d^{3}k}{(2\pi)^{3}2k} \lbrace \dot{\phi}^2 + \frac{1}{2} \partial_\lambda \phi \partial^\lambda \phi \rbrace 
\end{align} 
\begin{align}
3p =  N \int\frac{d^{3}k}{(2\pi)^{3}2k} \lbrace \partial_i \phi \partial^i \phi  - \frac{3}{2} \partial_\lambda \phi \partial^\lambda \phi \rbrace 
\end{align} 

Note that taking in the integral measure $\omega_k = k$ is just an adiabatic approximation but sufficient to study the second order contributions we are interested in.

As shown in \cite{Maggiore,Fulling1} inserting the mode expansion in this expression and evaluating it
between vacuum states with the condition $\langle0|a_{k'}a_{k}^{\dagger}|0\rangle=(2\pi)^{2}k\,\v(k-k')$
gives the vacuum energy density and pressure: 

\begin{align}
\r_{\text{vac}}=\frac{N}{2}\int\frac{d^{3}k}{(2\pi)^{3}2k}\left(|\dot{\phi}_{k}|^{2}+\frac{k^{2}}{a^{2}}|\phi_{k}|^{2}\right), \\
p_{\text{vac}}=\frac{N}{2}\int\frac{d^{3}k}{(2\pi)^{3}2k}\left(|\dot{\phi}_{k}|^{2}-\frac{k^{2}}{3a^{2}}|\phi_{k}|^{2}\right).
\end{align}

For the de Sitter solution using

\begin{align}
\label{eq:eom}
\dot{\phi}_{k}=\frac{1}{a}\phi'_{k}=-i\frac{k}{a^{2}}e^{ik/\dot{a}}\;\Rightarrow\;|\dot{\phi}_{k}|^{2}=\frac{k^{2}}{a^{4}} \\
\text{and}\quad|\phi_{k}|^{2}=\frac{1}{a^{2}}+\frac{H^{2}}{k^{2}},
\end{align}

one obtains: 

\begin{align}
\label{eq:rho}
\frac{\r_{\text{vac}}^{\text{FRW}}}{N}=\int\frac{d^{3}k}{(2\pi)^{3}2k}\left(\frac{k^{2}}{a^{4}}+\frac{H^{2}}{2a^{2}}\right) \\ \nonumber
=\int\frac{d^{3}p}{(2\pi)^{3}2p}\left(p^{2}+\frac{H^{2}}{2}\right),\\
\label{eq:p}
\frac{p_{\text{vac}}^{\text{FRW}}}{N}=\int\frac{d^{3}k}{(2\pi)^{3}2k}\,\frac{1}{3}\left(\frac{k^{2}}{a^{4}}-\frac{H^{2}}{2a^{2}}\right)\\ \nonumber
=\int\frac{d^{3}p}{(2\pi)^{3}2p}\,\frac{1}{3}\left(p^{2}-\frac{H^{2}}{2}\right).
\end{align}

Note that $k$ is just a mode label and that the physical momentum
is $p=k/a$, hence we made this change of variables to physical momenta. 
This implies for the energy density and pressure

\begin{align}
\label{LapseDefinition}
\r_{\text{vac}}=N\int\frac{d^{3}p}{(2\pi)^{3}2p}\left(\frac{H^{2}}{2}\right),\\
p_{\text{vac}}=-\frac{N}{3}\int\frac{d^{3}p}{(2\pi)^{3}2p}\,\left( \frac{H^{2}}{2}\right).
\end{align}
These values are in agreement with out normalization condition and the lapse function $N$ leaves the freedom of time reparametrizations.

There is an other way in which the expression for the vacuum energy can be formulated such that explicit spatial momentum dependencies vanish. Let us compute the vacuum expectation of the trace of the energy-momentum tensor and use \eqn (\ref{eq:eom}) as deSitter solutions for the field equations. 

\begin{align}
\label{eq:trace}
\langle0|T{}_{\m}^\m|0\rangle_{\text{vac}}=-N\langle0| \partial_\m\phi \partial^\m \phi  |0\rangle_{\text{vac}}=N\int\frac{d^{3}p}{(2\pi)^{3}2p}\,H^{2}.
\end{align}
The trace gives $\langle0|T{}_{\m}^\m|0\rangle_{\text{vac}}= \rho_{\text{vac}}- 3 p_{\text{vac}}$. But we obtained an integral for the trace which is divergent and needs to be regularized in order to make any physical statement. We will now turn to this issue.

\section{\label{sec:Regularization} Regularization}

The integrals we obtained for the vacuum energy and pressure are divergent, as we did not subtract the second adiabatic order.
Therefore, a regularization procedure is needed to extract sensible
values out of them. 

At first we consider the cut-off regulator which truncates the mode-sum. In flat space-time a covariant 4-momentum cut-off can be applied which is consistent with the symmetry of the theory.
We will demonstrate now that on curved space-time this is not the case.

\subsection{The 4-momentum cut-off}

That a Planck scale cut-off is related to the cosmological constant problem has been argued for in the literature , as in \cite{Maggiore} to extract a natural value for the vacuum energy. We introduce a momentum space cut-off $\Lambda$ which is assumed to be a Lorentz scalar and covariant in Minkowski space-time. We note that the Hubble parameter in our vacuum integrals \eqn (\ref{LapseDefinition}) does not depend on the momentum and thus 
we can use the relativistic  invariant form of the integral as as discussed in
\cite{Evgeny}

\begin{align}
\int\frac{d^{3}p}{(2\pi)^{3}2p}=i\int\frac{d^{4}p}{(2\pi)^{4}}\frac{1}{p^{2}+i\epsilon}.
\end{align}

Performing a Wick rotation and introducing $\Lambda$ as the 4-momentum
cut-off we obtain

\begin{align}
\int\frac{d^{4}p_{E}}{(2\pi)^{4}}\frac{1}{p^{2}}=\frac{1}{(4\pi)^{2}}\Lambda^{2}.
\end{align}

Thus, the vacuum energy has the value 

\begin{align}
\r_{\text{vac}}=\frac{N\,H^{2}\Lambda^{2}}{2(4\pi)^{2}}
\end{align}

and the pressure 

\begin{align}
p_{\text{vac}}=- \frac{1}{3}\frac{N\, H^{2}\Lambda^{2}}{2(4\pi)^{2}}.
\end{align}

The vacuum expectation for the trace of the energy-momentum tensor \eqn (\ref{eq:trace}) can be regularized in the same way and yields 
\begin{align}
\langle0|T{}_{\m}^\m|0\rangle_{\text{vac}}=-\frac{N\,H^{2}\Lambda^{2}}{(4\pi)^{2}}=-2\r_{\text{vac}}.
\end{align}

This implies, given our metric convention that $p_{\text{vac}}=-1/3 \rho_{\text{vac}}$, which is consistent with the direct calculation. We observe that even with a 4-momentum cut-off, which respects the remaining Lorentz symmetry, general covariance is broken. 

The issue of broken general covariance by the 3-momentum cut-off is discussed in \cite{Hollenstein:2011cz}, it is suggested to introduce non-covariant counter terms to remedy this problem. Those would compensate the breaking and the full theory is considered covariant again. This can be done analogously here for example 
adding a counterterm which results in an additional tensor $T_{\m\n}^c$, such that 

\begin{align}
\nabla^\m \left( T_{\m\n}^{\text{vac}} +  T_{\m\n}^c \right)=0.
\end{align}
We see that in our de Sitter set-up the  $\Lambda^2H^2 = \text{const}$ contribution has the covariant form $\Lambda^2 H^2\,g_{\mu\nu} \propto G_{\mu\nu}$. Since it is proportional to the Einstein Tensor it  renormalizes the Newton constant and is thus without physical consequences. At this point it is important to stress the difference between the gauge invariance of QFT and general covariance. As opposed to covariance under diffeomorphisms there is no physical gauge. However, it is possible that quantities depend on the frame as long as they transform correctly when the frame is changed. Furthermore, there are physically preferred frames whenever a material system is associated with it. If we would parametrize the time with a different lapse $\tilde{N}$, the relation between the $T_{\m\n}^{\text{vac}}$ and $ T_{\m\n}^c $ would change and needed to be readjusted by hand i.e. the counterterm choice would depend on the reference frame. 
If one wished to interpret this on physical grounds a mechanism would be needed to explain such behaviour. Examples of such completions are known to exist, see \cite{Green:1984sg}. However, since any convincing mechanism is missing this should be rather seen as a technical step and for sure no statement about "naturalness" of the actual value of the vacuum energy can be made.

\section{Renormalization}

We have seen that applying a regulator which does not respect the symmetries of the underlying theory leads to technical difficulties and requires adjustments of counterterms by hand. Thus it seems more easy to apply a regularization procedure which is consistent with the fundamental symmetries. There are methods to regularize vacuum fluctuations in a gravitational background which respect general covariance, as discussed in \cite{Birrell:1982ix, Lavrov:2009bv, Asorey:2012xq, Koksma:2011cq}. For a mass-less scalar field covariant calculations have only contributions of order $H^4$ to the vacuum energy, which our adiabatically limited expansion has missed. 

We would like to stress the following point.  As it has been argued by \cite{Bardeen:1995kv,Meissner:2006zh,Foot:2007ay,Foot:2007iy} that if classical scale symmetry is postulated the Higgs mass is protected against power divergences, given that there are no heavy fundamental scalar scales present in the theory. 

We demonstrated in this letter that general covariance has the same  protective effect when it comes to the cosmological constant, as that  of conformal symmetry  for the Higgs mass. However, we know that general covariance is a symmetry of nature and does not need to be postulated in addition. The only condition is that there is no hard cut-off scale connected to a finite physical length which would in fact break general covariance in the UV sector. Given that, the renormalization conditions needed to ensure general covariance allow only  vacuum contributions which are connected to breaking of scale invariance as known from conformal anomalies.

\section{\label{sec:Conclusion}Conclusion}

In this letter we discussed regularization of  vacuum fluctuations in a gravitational background and its dependence on the regularization procedure.  As well known a 3-momentum cut-off violates the Lorentz symmetry. This can be remedied by applying a 4-momentum cut-off in flat space-time. We show that in a general background covariance is violated even with the 4-momentum regulator. 

We argue that unless the cut-off is connected to a physical frame the terms violating covariance have to be augmented by frame dependent counterterms and absorbed in the Lagrangian quantities without observable effects. Thus it becomes obvious that general covariance acts in the same way as a protective symmetry for the cosmological constant as conformal symmetry is suggested to stabilize the Higgs mass.

\subsection*{Acknowledgements}

I cordially thank Evgeny Akhmedov, Roberto Cassadio, Julian Heeck, Alexander Kartavtsev, Manfred Lindner,  Krsitian McDonald and  Ilja Shapiro for helpful scientific discussions. I acknowledge the support by the IMPRS for Precision Tests of Fundamental Symmetries during this work.


\begin{thebibliography}{28}
\expandafter\ifx\csname natexlab\endcsname\relax\def\natexlab#1{#1}\fi
\expandafter\ifx\csname bibnamefont\endcsname\relax
  \def\bibnamefont#1{#1}\fi
\expandafter\ifx\csname bibfnamefont\endcsname\relax
  \def\bibfnamefont#1{#1}\fi
\expandafter\ifx\csname citenamefont\endcsname\relax
  \def\citenamefont#1{#1}\fi
\expandafter\ifx\csname url\endcsname\relax
  \def\url#1{\texttt{#1}}\fi
\expandafter\ifx\csname urlprefix\endcsname\relax\def\urlprefix{URL }\fi
\providecommand{\bibinfo}[2]{#2}
\providecommand{\eprint}[2][]{\url{#2}}

\bibitem[{\citenamefont{Prange}(1999)}]{Prange:1997iy}
\bibinfo{author}{\bibfnamefont{D.}~\bibnamefont{Prange}},
  \bibinfo{journal}{J.Phys.} \textbf{\bibinfo{volume}{A32}},
  \bibinfo{pages}{2225} (\bibinfo{year}{1999}), \eprint{hep-th/9710225}.

\bibitem[{\citenamefont{Brunetti and Fredenhagen}(2009)}]{Fredenhagen}
\bibinfo{author}{\bibfnamefont{R.}~\bibnamefont{Brunetti}} \bibnamefont{and}
  \bibinfo{author}{\bibfnamefont{K.}~\bibnamefont{Fredenhagen}}
  (\bibinfo{year}{2009}), \eprint{0901.2063}.

\bibitem[{\citenamefont{Maggiore}(2011)}]{Maggiore}
\bibinfo{author}{\bibfnamefont{M.}~\bibnamefont{Maggiore}},
  \bibinfo{journal}{Phys.Rev.} \textbf{\bibinfo{volume}{D83}},
  \bibinfo{pages}{063514} (\bibinfo{year}{2011}), \eprint{1004.1782}.

\bibitem[{\citenamefont{Bernard and LeClair}(2013)}]{Bernard:2012nv}
\bibinfo{author}{\bibfnamefont{D.}~\bibnamefont{Bernard}} \bibnamefont{and}
  \bibinfo{author}{\bibfnamefont{A.}~\bibnamefont{LeClair}},
  \bibinfo{journal}{Phys.Rev.} \textbf{\bibinfo{volume}{D87}},
  \bibinfo{pages}{063010} (\bibinfo{year}{2013}), \eprint{1211.4848}.

\bibitem[{\citenamefont{Casadio}(2013)}]{Casadio:2013uia}
\bibinfo{author}{\bibfnamefont{R.}~\bibnamefont{Casadio}},
  \bibinfo{journal}{Phys.Lett.} \textbf{\bibinfo{volume}{B724}},
  \bibinfo{pages}{351} (\bibinfo{year}{2013}), \eprint{1303.1914}.

\bibitem[{\citenamefont{Mangano}(2010)}]{Mangano:2010hw}
\bibinfo{author}{\bibfnamefont{G.}~\bibnamefont{Mangano}},
  \bibinfo{journal}{Phys.Rev.} \textbf{\bibinfo{volume}{D82}},
  \bibinfo{pages}{043519} (\bibinfo{year}{2010}), \eprint{1005.2758}.

\bibitem[{\citenamefont{Sushkov et~al.}(2011)\citenamefont{Sushkov, Kim,
  Dalvit, and Lamoreaux}}]{Lamoreaux}
\bibinfo{author}{\bibfnamefont{A.}~\bibnamefont{Sushkov}},
  \bibinfo{author}{\bibfnamefont{W.}~\bibnamefont{Kim}},
  \bibinfo{author}{\bibfnamefont{D.}~\bibnamefont{Dalvit}}, \bibnamefont{and}
  \bibinfo{author}{\bibfnamefont{S.}~\bibnamefont{Lamoreaux}},
  \bibinfo{journal}{Nature Phys.} \textbf{\bibinfo{volume}{7}},
  \bibinfo{pages}{230} (\bibinfo{year}{2011}), \eprint{1011.5219}.

\bibitem[{\citenamefont{{Wilson} et~al.}(2011)\citenamefont{{Wilson},
  {Johansson}, {Pourkabirian}, {Simoen}, {Johansson}, {Duty}, {Nori}, and
  {Delsing}}}]{DC}
\bibinfo{author}{\bibfnamefont{C.~M.} \bibnamefont{{Wilson}}},
  \bibinfo{author}{\bibfnamefont{G.}~\bibnamefont{{Johansson}}},
  \bibinfo{author}{\bibfnamefont{A.}~\bibnamefont{{Pourkabirian}}},
  \bibinfo{author}{\bibfnamefont{M.}~\bibnamefont{{Simoen}}},
  \bibinfo{author}{\bibfnamefont{J.~R.} \bibnamefont{{Johansson}}},
  \bibinfo{author}{\bibfnamefont{T.}~\bibnamefont{{Duty}}},
  \bibinfo{author}{\bibfnamefont{F.}~\bibnamefont{{Nori}}}, \bibnamefont{and}
  \bibinfo{author}{\bibfnamefont{P.}~\bibnamefont{{Delsing}}},
  \bibinfo{journal}{\nat} \textbf{\bibinfo{volume}{479}}, \bibinfo{pages}{376}
  (\bibinfo{year}{2011}), \eprint{1105.4714}.

\bibitem[{\citenamefont{Birrell and Davies}(1982)}]{Birrell:1982ix}
\bibinfo{author}{\bibfnamefont{N.}~\bibnamefont{Birrell}} \bibnamefont{and}
  \bibinfo{author}{\bibfnamefont{P.}~\bibnamefont{Davies}},
  \bibinfo{journal}{Cambridge Monogr.Math.Phys.}  (\bibinfo{year}{1982}).


\bibitem[{\citenamefont{Fulling et~al.}(1974)\citenamefont{Fulling, Parker, and
  Hu}}]{Fulling:1974pu}
\bibinfo{author}{\bibfnamefont{S.}~\bibnamefont{Fulling}},
  \bibinfo{author}{\bibfnamefont{L.}~\bibnamefont{Parker}}, \bibnamefont{and}
  \bibinfo{author}{\bibfnamefont{B.}~\bibnamefont{Hu}},
  \bibinfo{journal}{Phys.Rev.} \textbf{\bibinfo{volume}{D10}},
  \bibinfo{pages}{3905} (\bibinfo{year}{1974}).


\bibitem[{\citenamefont{Parker and Fulling}(1974)}]{Fulling1}
\bibinfo{author}{\bibfnamefont{L.}~\bibnamefont{Parker}} \bibnamefont{and}
  \bibinfo{author}{\bibfnamefont{S.}~\bibnamefont{Fulling}},
  \bibinfo{journal}{Phys.Rev.} \textbf{\bibinfo{volume}{D9}},
  \bibinfo{pages}{341} (\bibinfo{year}{1974}).

\bibitem[{\citenamefont{Dodonov and Klimov}(1996)}]{Dodonov}
\bibinfo{author}{\bibfnamefont{V.}~\bibnamefont{Dodonov}} \bibnamefont{and}
  \bibinfo{author}{\bibfnamefont{A.}~\bibnamefont{Klimov}},
  \bibinfo{journal}{Phys.Rev.} \textbf{\bibinfo{volume}{A53}},
  \bibinfo{pages}{2664} (\bibinfo{year}{1996}).

\bibitem[{\citenamefont{Arnowitt et~al.}(2008)\citenamefont{Arnowitt, Deser,
  and Misner}}]{ADM}
\bibinfo{author}{\bibfnamefont{R.~L.} \bibnamefont{Arnowitt}},
  \bibinfo{author}{\bibfnamefont{S.}~\bibnamefont{Deser}}, \bibnamefont{and}
  \bibinfo{author}{\bibfnamefont{C.~W.} \bibnamefont{Misner}},
  \bibinfo{journal}{Gen.Rel.Grav.} \textbf{\bibinfo{volume}{40}},
  \bibinfo{pages}{1997} (\bibinfo{year}{2008}), \eprint{gr-qc/0405109}.

\bibitem[{\citenamefont{Hollenstein et~al.}(2012)\citenamefont{Hollenstein,
  Jaccard, Maggiore, and Mitsou}}]{Hollenstein:2011cz}
\bibinfo{author}{\bibfnamefont{L.}~\bibnamefont{Hollenstein}},
  \bibinfo{author}{\bibfnamefont{M.}~\bibnamefont{Jaccard}},
  \bibinfo{author}{\bibfnamefont{M.}~\bibnamefont{Maggiore}}, \bibnamefont{and}
  \bibinfo{author}{\bibfnamefont{E.}~\bibnamefont{Mitsou}},
  \bibinfo{journal}{Phys.Rev.} \textbf{\bibinfo{volume}{D85}},
  \bibinfo{pages}{124031} (\bibinfo{year}{2012}), \eprint{1111.5575}.

\bibitem[{\citenamefont{Dvali et~al.}(2007)\citenamefont{Dvali, Hofmann, and
  Khoury}}]{Dvali}
\bibinfo{author}{\bibfnamefont{G.}~\bibnamefont{Dvali}},
  \bibinfo{author}{\bibfnamefont{S.}~\bibnamefont{Hofmann}}, \bibnamefont{and}
  \bibinfo{author}{\bibfnamefont{J.}~\bibnamefont{Khoury}},
  \bibinfo{journal}{Phys.Rev.} \textbf{\bibinfo{volume}{D76}},
  \bibinfo{pages}{084006} (\bibinfo{year}{2007}), \eprint{hep-th/0703027}.

\bibitem[{\citenamefont{Akhmedov}(2002)}]{Evgeny}
\bibinfo{author}{\bibfnamefont{E.~K.} \bibnamefont{Akhmedov}}
  (\bibinfo{year}{2002}), \eprint{hep-th/0204048}.

\bibitem[{\citenamefont{Consoli and Costanzo}(2008)}]{Consoli:2007cw}
\bibinfo{author}{\bibfnamefont{M.}~\bibnamefont{Consoli}} \bibnamefont{and}
  \bibinfo{author}{\bibfnamefont{E.}~\bibnamefont{Costanzo}},
  \bibinfo{journal}{Eur.Phys.J.} \textbf{\bibinfo{volume}{C54}},
  \bibinfo{pages}{285} (\bibinfo{year}{2008}), \eprint{0709.4101}.

\bibitem[{\citenamefont{Lavrov and Shapiro}(2010)}]{Lavrov:2009bv}
\bibinfo{author}{\bibfnamefont{P.~M.} \bibnamefont{Lavrov}} \bibnamefont{and}
  \bibinfo{author}{\bibfnamefont{I.~L.} \bibnamefont{Shapiro}},
  \bibinfo{journal}{Phys.Rev.} \textbf{\bibinfo{volume}{D81}},
  \bibinfo{pages}{044026} (\bibinfo{year}{2010}), \eprint{0911.4579}.

\bibitem[{\citenamefont{Asorey et~al.}(2012)\citenamefont{Asorey, Lavrov,
  Ribeiro, and Shapiro}}]{Asorey:2012xq}
\bibinfo{author}{\bibfnamefont{M.}~\bibnamefont{Asorey}},
  \bibinfo{author}{\bibfnamefont{P.~M.} \bibnamefont{Lavrov}},
  \bibinfo{author}{\bibfnamefont{B.~J.} \bibnamefont{Ribeiro}},
  \bibnamefont{and} \bibinfo{author}{\bibfnamefont{I.~L.}
  \bibnamefont{Shapiro}}, \bibinfo{journal}{Phys.Rev.}
  \textbf{\bibinfo{volume}{D85}}, \bibinfo{pages}{104001}
  (\bibinfo{year}{2012}), \eprint{1202.4235}.

\bibitem[{\citenamefont{Koksma and Prokopec}(2011)}]{Koksma:2011cq}
\bibinfo{author}{\bibfnamefont{J.~F.} \bibnamefont{Koksma}} \bibnamefont{and}
  \bibinfo{author}{\bibfnamefont{T.}~\bibnamefont{Prokopec}}
  (\bibinfo{year}{2011}), \eprint{1105.6296}.

\bibitem[{\citenamefont{Green and Schwarz}(1984)}]{Green:1984sg}
\bibinfo{author}{\bibfnamefont{M.~B.} \bibnamefont{Green}} \bibnamefont{and}
  \bibinfo{author}{\bibfnamefont{J.~H.} \bibnamefont{Schwarz}},
  \bibinfo{journal}{Phys.Lett.} \textbf{\bibinfo{volume}{B149}},
  \bibinfo{pages}{117} (\bibinfo{year}{1984}).

\bibitem[{\citenamefont{Martin}(2012)}]{Martin:2012bt}
\bibinfo{author}{\bibfnamefont{J.}~\bibnamefont{Martin}},
  \bibinfo{journal}{Comptes Rendus Physique} \textbf{\bibinfo{volume}{13}},
  \bibinfo{pages}{566} (\bibinfo{year}{2012}), \eprint{1205.3365}.

\bibitem[{\citenamefont{Bardeen}(1995)}]{Bardeen:1995kv}
\bibinfo{author}{\bibfnamefont{W.~A.} \bibnamefont{Bardeen}}
  (\bibinfo{year}{1995}).

\bibitem[{\citenamefont{Meissner and Nicolai}(2007)}]{Meissner:2006zh}
\bibinfo{author}{\bibfnamefont{K.~A.} \bibnamefont{Meissner}} \bibnamefont{and}
  \bibinfo{author}{\bibfnamefont{H.}~\bibnamefont{Nicolai}},
  \bibinfo{journal}{Phys.Lett.} \textbf{\bibinfo{volume}{B648}},
  \bibinfo{pages}{312} (\bibinfo{year}{2007}), \eprint{hep-th/0612165}.

\bibitem[{\citenamefont{Foot et~al.}(2007)\citenamefont{Foot, Kobakhidze,
  McDonald, and Volkas}}]{Foot:2007ay}
\bibinfo{author}{\bibfnamefont{R.}~\bibnamefont{Foot}},
  \bibinfo{author}{\bibfnamefont{A.}~\bibnamefont{Kobakhidze}},
  \bibinfo{author}{\bibfnamefont{K.}~\bibnamefont{McDonald}}, \bibnamefont{and}
  \bibinfo{author}{\bibfnamefont{R.}~\bibnamefont{Volkas}},
  \bibinfo{journal}{Phys.Rev.} \textbf{\bibinfo{volume}{D76}},
  \bibinfo{pages}{075014} (\bibinfo{year}{2007}), \eprint{0706.1829}.

\bibitem[{\citenamefont{Foot et~al.}(2008)\citenamefont{Foot, Kobakhidze,
  McDonald, and Volkas}}]{Foot:2007iy}
\bibinfo{author}{\bibfnamefont{R.}~\bibnamefont{Foot}},
  \bibinfo{author}{\bibfnamefont{A.}~\bibnamefont{Kobakhidze}},
  \bibinfo{author}{\bibfnamefont{K.~L.} \bibnamefont{McDonald}},
  \bibnamefont{and} \bibinfo{author}{\bibfnamefont{R.~R.}
  \bibnamefont{Volkas}}, \bibinfo{journal}{Phys.Rev.}
  \textbf{\bibinfo{volume}{D77}}, \bibinfo{pages}{035006}
  (\bibinfo{year}{2008}), \eprint{0709.2750}.

\bibitem[{\citenamefont{Foot et~al.}(2011)\citenamefont{Foot, Kobakhidze, and
  Volkas}}]{Foot:2010et}
\bibinfo{author}{\bibfnamefont{R.}~\bibnamefont{Foot}},
  \bibinfo{author}{\bibfnamefont{A.}~\bibnamefont{Kobakhidze}},
  \bibnamefont{and} \bibinfo{author}{\bibfnamefont{R.~R.}
  \bibnamefont{Volkas}}, \bibinfo{journal}{Phys.Rev.}
  \textbf{\bibinfo{volume}{D84}}, \bibinfo{pages}{075010}
  (\bibinfo{year}{2011}), \eprint{1012.4848}.

\bibitem[{\citenamefont{Shapiro and Sola}(2009)}]{Shapiro:2009dh}
\bibinfo{author}{\bibfnamefont{I.~L.} \bibnamefont{Shapiro}} \bibnamefont{and}
  \bibinfo{author}{\bibfnamefont{J.}~\bibnamefont{Sola}},
  \bibinfo{journal}{Phys.Lett.} \textbf{\bibinfo{volume}{B682}},
  \bibinfo{pages}{105} (\bibinfo{year}{2009}), \eprint{0910.4925}.

\bibitem[{\citenamefont{Gibbons}(2014)}]{Gibbons:2014zla}
\bibinfo{author}{\bibfnamefont{G.}~\bibnamefont{Gibbons}}
  (\bibinfo{year}{2014}), \eprint{1403.5431}.

\end{thebibliography}

\end{document}